\documentclass[aps,prl,reprint,groupedaddress,nofootinbib]{revtex4-1}

\usepackage{graphicx}
\usepackage{longtable}

\newcommand{\gt}{\tilde g}

\newcommand{\xc}{\tilde\chi^0_1}

\newcommand{\xt}{\tilde t_1}

\usepackage{cancel}
 \usepackage[normalem]{ulem}
\usepackage{color}

\begin{document}

\title{The LHC Diphoton excess at 750 GeV in the framework 
of the Constrained Minimal Supersymmetric Standard Model}

\author{\vskip -10pt Debajyoti Choudhury$^1$ and Kirtiman Ghosh$^2$}
\affiliation{Department of Physics and Astrophysics, University of Delhi, Delhi 110007, India\\
$^1$debajyoti.choudhury@gmail.com, $^2$kirti.gh@gmail.com}


\begin{abstract}
We examine the observed diphoton excess at 750 GeV in the framework of
constrained Minimal Supersymmetric Standard Model (cMSSM). The
consistency of cMSSM with the discovery of a 125 GeV Higgs boson and
PLANCK/WMAP data for dark matter relic density (RD) demands large
mixing in the scalar top (stop) sector and consequently, the existence
of a light stop quasi-degenerate with the lightest neutralino. Small
decay width due to kinematic and loop suppression of such a stop
allows for the existence of stop-antistop bound states (stoponia).
Identifying the particular part of cMSSM parameter space consistent
with Higgs boson mass as well as RD, we study stoponia as the source
of the diphoton excess.
\end{abstract}
\maketitle

Recent LHC reports \cite{atlas_di,cms_di} of a diphoton excess at about
750 GeV invariant mass could be interpreted as production and decay of
a new massive spin-0 or spin-2 particle. The ATLAS collaboration
observed the most significant deviation from the background
predictions at $M_{\gamma \gamma}\sim$750 GeV with a local
significance of 3.6$\sigma$ and 3.9$\sigma$ in searches optimized for
a spin-2 and spin-0 particle, respectively. For the CMS collaboration,
the largest excess is observed around $M_{\gamma \gamma}\sim760$ GeV
with corresponding local significance of 2.8$\sigma$
(2.9$\sigma$). Under the narrow width approximation, the fitted LHC 13
TeV ATLAS and CMS data correspond to a resonance with effective signal
strength (namely, production cross-section times branching ratio into
$\gamma \gamma$) of $10\pm 3$ and $2 \sim 6$ fb, respectively.  
The best fitted values for the decay width are 45 GeV for the
ATLAS data and 0.1 GeV for CMS Run I and II data
combined. This large discrepancy is perhaps reflective of the limited 
statistics and we shall return to this issue later.

The excess has motivated many phenomenological studies
\cite{Ellis:2015oso,generic_susy}
of possible scenarios of new physics beyond the SM.  The mutual
consistency of the LHC 8 TeV (no diphoton signature)
\cite{Aad:2015mna,CMS:2014onr} and 13 TeV (diphoton excess) results
demands that gluon-gluon fusion be the dominant mechanism for the
production of the resonance. Adopting the spin-0
hypothesis\footnote{While a Randall-Sundrum-type graviton has been
offered as a solution, suppressing the decay into a $t\bar t$ or a
dilepton pair (none of which has been seen), is not very
straightforward \cite{Arun:2015ubr}.},
the (pseudo)scalar couples with a pair of gluons/photons only at the
loop level. To explain the signal size and profile, one needs to
postulate additional (and heavy enough not to have been seen) colored
and charged states with large couplings to the resonance so as to
enhance its induced coupling to a pair of gluons/photons.  On the
other hand, if the resonance is a QCD bound-state of new colored
(charged) particles then it automatically couples to a pair of gluons
(photons) via the annihilation Feynman diagram. Resonant production of
particle-antiparticle ($X\bar X$) bound-state ($\eta_X$) at collider
is possible if the formation time ($\sim$ the inverse of binding
energy, $E_b$) of $\eta_X$ is smaller than the life-time ($\sim$
inverse of decay width, $\Gamma_X$) of $X$: $\Gamma_X \ll E_b$.
Subsequently, the decay of the $\eta_X$ may proceed either through the
prompt decay of the constituent $X$ or through annihilation
diagrams. If the annihilation decay width ($\Gamma_{\eta_X}$)
dominates over $\Gamma_X$, we could, potentially, observe the resonant
signature $\eta_X$ at colliders before discovering the new particle
$X$. Historically, the first signals for c-- and b-quarks in hadron
colliders were leptonic decays of their $J/\psi(\bar c c)$ or
$\Upsilon(\bar b b)$ bound states. Therefore, if there is a colored
particle in nature with few hundred GeV mass and suppressed decay,
then the signature of the bound state could pop up before the
discovery of the particle itself. As a result, there is a fair amount
of effort going on to explain the 750 GeV diphoton resonance as a QCD
bound state
\cite{Hamaguchi:2016umx}.
In this work, we explain the diphoton excess in the framework of
supersymmetry (SUSY), which predicts the existence of TeV scale
colored fermions (partners of gluons namely, gluinos) and scalars
(partners of the SM quarks namely, squarks). To retain a viable dark
matter (DM) candidate, we will impose $R$-parity
conservation. Consequently, if the tree-level 2-body decay of a
colored SUSY-particle into a lighter SUSY-particle is kinematically
(or otherwise) suppressed, then its individual decay width could be
small enough to give rise to a bound-state, albeit short-lived.  In
particular, if the mass splitting between the of top-squark (the stop)
and the lightest neutralino $\xc$ (the lightest supersymmetric
particle or LSP) is small enough to forbid the 2-body decay of stop
into $t\xc$ (or a bottom-chargino pair), then a stop-antistop bound
pair (stoponium) is a distinct
possibility~\cite{stopon_generic,Martin:2008sv,Younkin:2009zn}
We explore this, restricting ourselves to the framework of the
constrained Minimal Supersymmetric Standard Model (cMSSM). 

Apart from usual motivations like stability of the Higgs mass, an
explanation for dark matter etc., the cMSSM has the advantage of
having a mathematically well constructed mechanism for supersymmetry
breaking at GUT scale.  With the phenomenology determined by only four
parameters at GUT scale namely, common SUSY breaking scalar ($m_0$)
and gaugino ($m_{1/2}$) masses, a single tri-linear mass parameter
$A_0$, ratio of Higgs vacuum expectation values ${\rm tan}\beta$ and
the discrete choice of the sign of the Higgs mixing parameter $\mu$,
the model is highly predictive and easy to probe in collider
experiments.  At the LHC, production of squarks and gluinos play the
most important roles in discovering cMSSM, with perhaps the best
signature being the multi-jets plus missing transverse energy
(${\cancel E}_T$) channel where ${\cancel E}_T$ results from the
weakly interacting LSPs appearing in the final state. However, at the
LHC RUN I and RUN II \cite{atlas_cmssm, cms_cmssm}, no such signature
has been detected and thus, equal squark and gluino masses below 1.85
TeV are excluded at 95\% confidence level (C.L.). These lower bounds
do not militate against the mechanism for stabilizing the electroweak
hierarchy from large radiative corrections, as only the third
generation squarks are important at one loop order and LHC searches
are not yet very sensitive for these.  Indeed, a stop mass of a few
hundred GeVs is still allowed and, in fact, favored by fine-tuning
arguments. Within the framework of cMSSM, a light stop can naturally
arises as a consequence of large third-generation Yukawa
coupling. With the off-diagonal terms in the mass matrix for the
squark of a given flavor being proportional to the Yukawa coupling and
growing with $A_0$, a large $A_0$ can generate a large mass splitting
in the stop sector and consequently produce a light stop ($\xt$) as
the next-to-lightest supersymmetric particle (NLSP).

 \begin{figure}[t]
 \includegraphics[angle=0,width=0.5\textwidth]{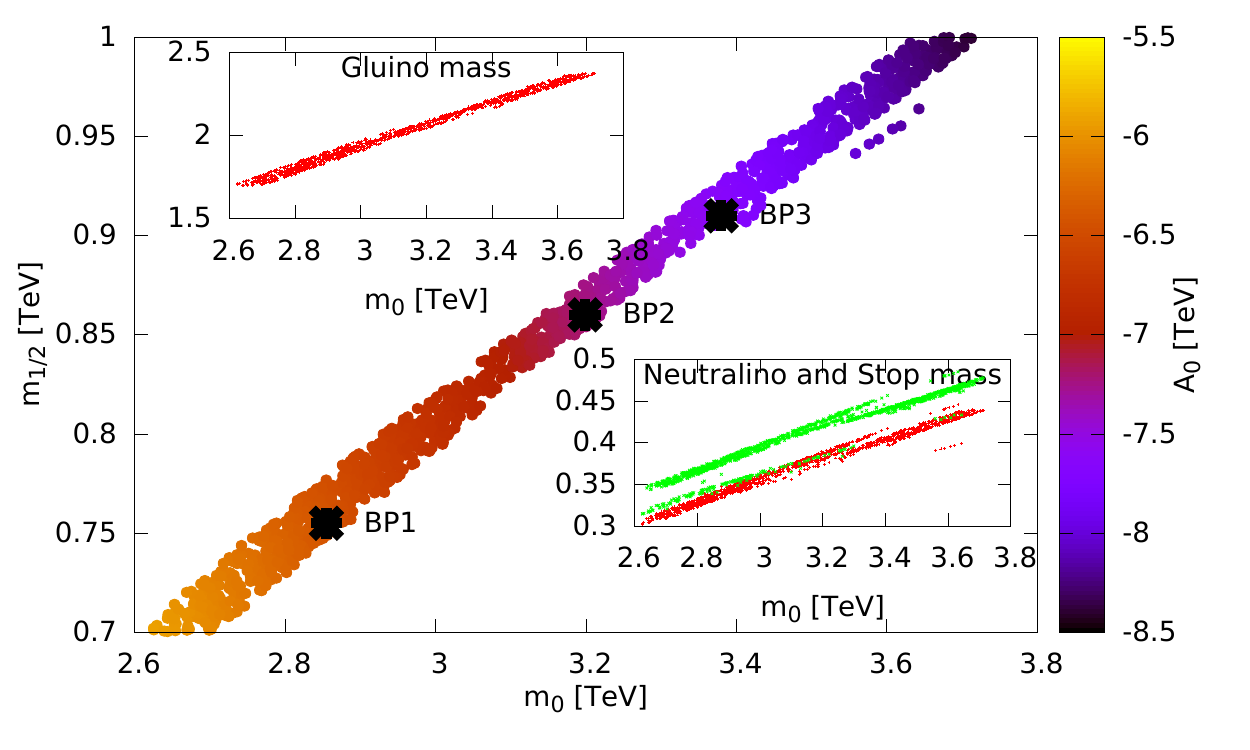}
 \caption{The 3-dimensional parameter space consistent with Higgs discovery and RD data, for ${\rm tan}\beta=15$ and $\mu>0$. The inset figures corresponds to scatter plots for LSP, lightest stop (bottom) and gluino (top) mass.}
\label{param_space}
\vskip -2pt
 \end{figure}

In fact, a quasi-degenerate $\xt$-NLSP is inherent to the only part of
the cMSSM parameter space which simultaneously explains WMAP/PLANCK
\cite{wmap,Ade:2015xua} DM relic density (RD) data as well as the LHC
discovery of a 126 GeV Higgs boson. Most of the cMSSM parameter space
gives rise to a bino-like LSP and, hence, too large a RD.  For a
quasi-degenerate $\xt$-$\xc$ pair, though, the observed relic density
can be explained via $\xt$-$\xc$ coannihilation. Furthermore,
electroweak (EW) baryogenesis favors a light-$\xt$ \cite{EWB}.
Interestingly, this very same part of the cMSSM parameter space
is also suitable for the production of stoponia at the LHC, and
we now carefully examine it.

With the off-diagonal term being given by
$M_{LR}^{2}=m_{t}(A_t-\mu{\rm cot}\beta)$, the mass splitting is
maximized when $\mu$ and $A_t$ have opposite signs.
 \begin{table}
\vskip -15pt
 \caption{Benchmark parameters and relevant sparticle masses in GeV (${\rm tan}\beta=15$, $\mu>0$).}
\label{BPs}
 \begin{ruledtabular}
\begin{tabular}{||c||c|c|c||c|c|c||c|c|c||}
BPs & $m_0$ & $m_{1/2}$ & $A_0$ & $m_{\gt}$ & $m_{\tilde t_2}$ & $m_{\xt}$ & $m_{\tilde \chi_{1}^{\pm}}$ & $m_{\xc}$ & $m_h$ \\\hline
BP1 & 2855 & 755 & -6467 & 1821 & 2213 & 370 & 643 & 334 & 127.3 \\ 
BP2 & 3199 & 860 & -7277 & 2052 & 2480 & 420 & 734 & 383 & 126.3 \\
BP3 & 3380 & 910 & -7695 & 2162 & 2616 & 445 & 777 & 406 & 127.5 \\
 \end{tabular}
 \end{ruledtabular}
\vskip -15pt
 \end{table}
Choosing $\mu>0$ and a representative value of $\tan\beta=15$, we scan
over $m_0$, $m_{1/2}$ and $A_0$ to obtain the phenomenologically
acceptable part of the parameter space.  The particle spectrum was
calculated using {\bf SuSpect} \cite{suspect} with $m_t=173.2$ GeV,
$m_b=4.2$ GeV, $m_{\tau}=1.777$ GeV and
$\alpha_s(m_Z)=0.1172$. Fig.~\ref{param_space} shows a 3-dimensional
scatter plot, in the $m_0$-$m_{1/2}$ plane with $A_0$ being indicated
by color gradient, of the parameter space consistent with the Higgs
mass $126 \pm 2$ GeV and WMAP/PLANCK RD data.  The density of cold
dark matter in the universe is determined to be \cite{wmap}
$\Omega_ch^2=0.1186\pm0.0036$. 
However, we only demand that the RD due
to $\xc$ (which we calculate using {\bf micrOmegas} \cite{microomega})
does not exceed the bound at the $2\sigma$ level. Furthermore, to
account for the higher order effects in RD calculations and the
consequent small inaccuracies in {\bf micrOmegas} \cite{Baro:2007em},
we relax the bound to $\Omega_ch^2<0.1311$. Fig.~\ref{param_space}
shows that there is only a narrow 
region in the cMSSM parameter space 
consistent with both Higgs mass and RD data. The situation is very
similar for other values of $\tan\beta$. In the insets of
Fig.~\ref{param_space}, we present the resulting $\xc,~\xt$ (bottom)
and $\gt$ (top) masses. In particular, note the small mass splitting
between the first two.

It is also important to check the
consistency of the model with other constraints like $BR(B\to X_s
\gamma)=(355\pm 142)\times 10^{-6}$ \cite{bsg}, $BR(B_s\to \mu^+
\mu^-) < 4.5 \times 10^{-9}$ (95\% C.L.) \cite{bmm},
$0.99<R^{NP}_{\tau
\nu_{\tau}}=\frac{BR(B^+\to\tau^+\nu_{\tau})_{SM+NP}}{BR(B^+\to\tau^+\nu_{\tau})_{SM}}<3.19$
\cite{btn} as well as direct collider searches at the LHC RUN I and
II. 
To illustrate further numerical results, we choose three benchmark
points (BP),  indicated by black dots in
Fig.~\ref{param_space} and listed in Table ~\ref{BPs}, 
along with  the masses of the relevant
sparticles and the light Higgs boson.
For the parameter-space in
Fig.~\ref{param_space}, the other squarks are heavier than the 
$\gt$ and, hence,
$\xc$, $\xt$ and $\gt$ are the most important
particles in the context of LHC searches. 
 In Table~\ref{cons}, we 
also list the next-to-leading (NLO)
order production cross-sections for $\xt \xt^*$ and $\gt \gt$ pairs at
the LHC with 8 and 13 TeV center-of-mass energy,  as
calculated using {\bf Prospino} 2.1 \cite{prospino} with CTEQ6.6M
\cite{cteq6.6m} parton distribution functions. Inspite of the 
substantial
production cross-section, $\xt$-signatures are hard to detect because
of the small $m_{\xt}$-$m_{\xc}$ splitting. The kinematically allowed
$\xt$ decays namely, loop induced flavor violating 2-body decay into
$c \xc$ and tree-level four body decay into a $b$-quark in association
with pair of light quarks and $\xc$, give rise to very soft jets and
${\cancel p}_T$ signature which is very hard to detect over the huge
SM backgrounds. Several alternative signatures like stop pair
production in association with two $b$-jets \cite{b_stop}, single
photon \cite{photon_stop} or jet \cite{jet_stop}, have been proposed
in the literature. Mono-jet or a charm-jet+${\cancel p}_T$ search at
the LHC with $\sqrt s=8$ TeV and ${\cal L}=20.3$ fb$^{-1}$ puts a
lower bound of 270 GeV \cite{LHC_lstop} on $\xt$-mass for a degenerate
$m_{\xt}$-$m_{\xc}$ scenario. 
As for $\gt \gt$-production, since the gluino overwhelmingly decays 
into a top-stop pair, the signature would be pair of tops accompanied by some 
very soft jets and ${\cancel p}_T$.
 For $\sqrt s=8$
TeV, the ATLAS \cite{LHC_hstop} collaboration studied an 
analogous process, namely stop-pair production with the
 stop decaying promptly
into $t\xc$-pairs. 
The negative result
can be roughly translated to a model independent upper bound of 13.9
fb on the BSM contribution to $t\bar t+{\cancel p}_T$. Table~\ref{cons}
shows that the gluino-pair production cross-sections are much smaller
than this bound.  Therefore, the BPs as well as the parameter space are
completely consistent with all the constraints.

 \begin{table}
 \caption{Low energy constraints, light stop decay width and NLO $\xt xt^*$, $\gt \gt$ production cross-section at the LHC 8 and 13 TeV.}
\label{cons}
 \begin{ruledtabular}
\begin{tabular}{||c||c|c|c||c||c|c||c|c||}
BPs & BR$^B_{X_s \gamma}$ & BR$^{B_s}_{\mu \mu}$ & $R^{NP}_{\tau \nu_{\tau}}$ & $\Omega_ch^2$ &\multicolumn{2}{|c||}{$\sigma(\xt \xt^*)$ [pb]}  & \multicolumn{2}{c||}{$\sigma(\gt \gt)$ [fb]} \\\cline{6-9}
    & $\times 10^{4}$ & $\times 10^{9}$ &  & eV & 8 & 13 & 8 & 13 \\\hline  
BP1 & 3.10 & 3.12 &  0.99 & 0.116 & 0.55 & 2.64 & 0.016 &  1.55  \\ 
BP2 & 3.14 & 3.11 & 0.99 & 0.122  & 0.25 & 1.32 & 0.002  &  0.45  \\
BP3 & 3.15 & 3.11 & 0.99 & 0.128  & 0.18 & 0.96 & 0.001 &  0.26  \\
 \end{tabular}
 \end{ruledtabular}
\vskip -15pt
 \end{table}

A stoponium can form only if its binding energy is greater than the
stop decay width. In the present case, apart from the phase space
suppression, the 2-body decay $\xt\to c\xc$ is also suppressed by the
loop as well as flavor violation factors, whereas the 4-body
decay $\xt \to b q \bar q^\prime \xc$ is suppressed due to the heavy
particles in the propagators.
$\xt$ decay widths for BP1, BP2 and BP3
are estimated (using SUSY-HIT1.5 \cite{susy_hit}) to be $1.28\times
10^{-10}$, $9.96\times 10^{-10}$ and $9.03\times 10^{-10}$ GeV
respectively. On the other hand, the typical binding energy of
a QCD bound-state is $E_b \sim \alpha_s m_{\xt}/a_0^2\sim {\cal
O}({\rm GeV})$ where $a_0$ is the Bohr radius of the system.  Thus,
stoponia may indeed exist. Once the bound-state is formed, it can
decay either via annihilation of its constituents to boson/fermion
pairs or via the decay of the constituents themselves.  Only the
former would give rise to a distinctive resonance signature.  The
annihilation partial decay widths of the $1S$ state into a pair of
gluons and photons are given by,
\begin{equation}
\Gamma_{\eta_{\xt}}^{gg}=\frac{4}{3}\, \alpha_S^{2}\, \Delta\qquad 
{\rm and} \qquad \Gamma_{\eta_{\xt}}^{\gamma\gamma}=\frac{32}{27}\, \alpha^2 
   \, \Delta \ ,
\label{decay}
\end{equation} 
where $\Delta$ is a parameter with a dimension of mass and, for a
non-relativistic (i.e., binding energy much smaller than the constituent
mass) bound-state, $\Delta$ is given in terms of 
the radial wavefunction at the origin 
$\psi(0) = \sqrt{4 \pi} R(0)$, viz.
$\Delta=|R(0)|^2/m_{\eta_{\xt}}^2$. In
Ref.~\cite{potential_model}, 
$|R(0)|$, and the 
binding energy of a non-relativistic bound state was computed
in the framework of potential models.
 Using $\Lambda_{\bar
{MS}}^{(4)}=300$ MeV 
and the parametrization of Ref~\cite{potential_model},
the non-relativistic stoponium binding energy (namely, $2m_{\xt}-m_{\eta_{\xt}}
\sim 3.274+1.777L+0.560L^2+0.081L^3$ GeV where $L=\ln(m_{\xt}/250{\rm
GeV})$) and stoponium mass for BP1 
are estimated to be 4.1 GeV and 736
GeV, respectively. Thus, BP1 could 
potentially explain the diphoton excess provided the stoponium production
cross-section and decay to $\gamma \gamma$ are consistent with the
observed excess. 

A recent study of the stoponium~\cite{NREFT} 
using the lattice formulation of
nonrelativistic effective field theory (NREFT),  
yields $\Delta\sim 0.57$ GeV for the 750 GeV 1S state, which is 
a factor of 3.5--4 larger than  that obtained within naive 
potential models. The consequent
partial decay width of 
such as state
into a pair of gluons 
(photons) can be
estimated to be $7.5\times 10^{-3}$ ($4.2\times 10^{-5}$) GeV. 
With the bound state decaying much faster than the constituents 
themselves, a resonant structure to the signal, thus, follows.

Stoponium production cross-section at the
LHC resulting from gluon-gluon fusion can be 
obtained by convoluting $\Gamma_{\eta_{\xt}}^{gg}$ with
the gluon densities, $g(x,Q^2)$, in the colliding protons:
\begin{equation}
\sigma^{pp}_{\eta_{\xt}}=\frac{\pi^2}{8m^3_{\eta_{\xt}}}\Gamma_{\eta_{\xt}}^{gg}\int_\tau^1 dx \frac{\tau}{x}g\left(x,Q^2\right)g\left(\frac{\tau}{x},Q^2\right),
\label{cross}
\end{equation}       
 where $\tau=m^2_{\eta_{\xt}}/s$.  Using the aforementioned widths, and
assuming that no other decay mode exists, the diphoton cross section
for $\sqrt s =13$ TeV, is approximately 0.19 fb. 
 In other words, within the 
the NREFT framework, the cross-section is nearly one order
of magnitude smaller than the observed excess. A few enhancements
are operative, though. For one, the NLO QCD correction
\cite{Younkin:2009zn} leads to a $K$-factor of approximately 1.38. 
Moreover, the 1S state is not the only one that is present or 
can be produced. For all the $nS$ states, the formalism is identical. 
However, with $R(0)$ being smaller, the individual 
production rates would be progressively smaller. And while their 
partial widths into the exclusive $\gamma\gamma$ mode are smaller, 
it is not so for the branching fraction, especially where the semi-inclusive 
rate (alongwith soft radiations) is concerned. As for the $P$-wave states, 
understandably, they cannot be formed directly from $gg$ fusion. However, 
the radiation of a soft gluon clearly allows this, much like the case 
of $J/\psi$ or the $\Upsilon$. However, even on the inclusion of all 
such effects, the overall enhancement is not expected to be larger than 
about 2.5. Thus, the observed rates cannot still be explained in totality.
More interestingly though, the large width observed by ATLAS could be 
partially explained owing to the spread in the masses of the bound states. 

 \begin{figure}[t]
\includegraphics[angle=-0,width=0.5\textwidth]{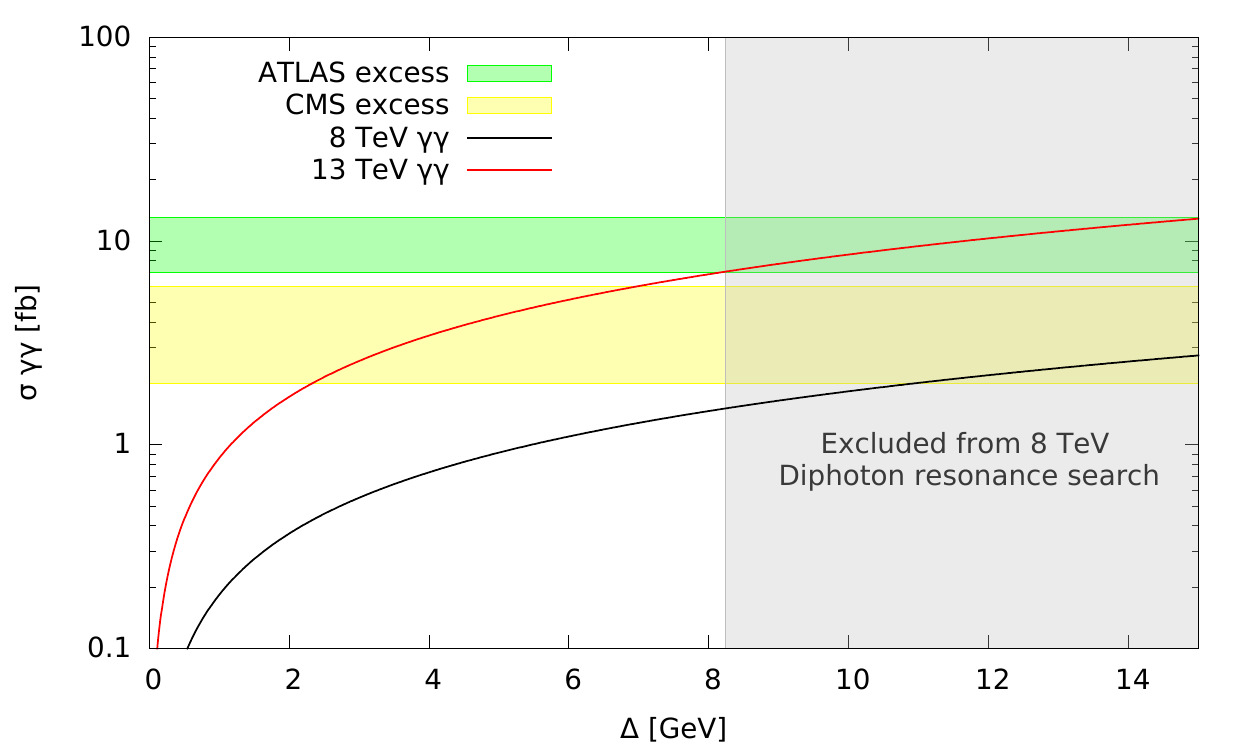}
 \caption{Production cross-section of a 750 GeV relativistic stoponium decaying into diphoton as a function of $\Delta$ at the LHC with 8 and 13 TeV center-of-mass energy. Yellow and green shaded regions indicate the fitted ranges for the diphoton excess at the CMS and ATLAS respectively.}
\label{sig}
\vskip -2pt
 \end{figure}

The large coupling $A_tH_u\tilde Q_L \tilde t_R$ provides another
attractive force between stop-antistop pairs mediated by the exchange
of the scalar quanta.  Analytically solving a Bethe-Salpeter equation
for the bound state in the framework of the Wick-Cutkosky model
\cite{WC} for massless scalar exchange, Ref.~\cite{Gian} predicted the
existence of two-stop composite states with binding energies of the
order of the electroweak scale for large $A_t$($\sim $ few TeV) and
small $m_{\xt}$($\sim $ few hundred GeV). The Wick-Cutkosky model for
a massive exchange scalar was studied in Ref.~\cite{gian7} and the
existence of relativistic bound-states was predicted for large enough
coupling.  There are several interesting consequences of such a
relativistic bound state.  For example, the stoponium could develop a
non-zero VEV and contribute to electroweak breaking
\cite{Gian}. Mixing of the stoponium with the Higgs could not only
generate EW-symmetry breaking from supersymmetry
breaking~\cite{peccei,peccei1}, but also enhance the Higgs
mass\footnote{Such considerations also save the scenario from
    potential constraints from color-breaking minima that a large
    $A_0$ often entails. With the inclusion of the bound states now
    being mandatory, the nominal one-loop corrected scalar potential
    \cite{Camargo-Molina:2013sta} involving only sfermions and Higgses
    does not hold any longer and the phase-diagram needs to be
    redetermined. While this also has a bearing on
    Fig.\ref{param_space}, there is no qualitative change.}.  The
critical value of $A_t$, required for symmetry-breaking seesaw to
occur, was estimated to be of the order of a few TeVs.

With the mass of a relativistic bound-state not being
well-understood in terms of the constituent mass (except for the fact
that the binding energy is relatively large), BP2 and BP3 are
representative points in the parameter space. Similarly,
$\Delta=|R(0)|^2/m_{\eta_{\xt}}^2$ is no longer applicable and
$\Delta$ has to be treated as an unknown parameter which is expected
to be much larger.  As Fig.~\ref{sig} shows, the diphoton excess can
be explained by relativistic stoponium production for $2.32~{\rm
GeV}<\Delta < 8.24~{\rm GeV}$. Once again, we have neglected both NLO
corrections as well as contributions from the excited
stoponia. Inclusion of these would move the $\Delta$-window to slightly
lower
values, while broadening the observed width.  
 
Some issues remain. We have, until now, neglected branching modes
other than than $gg$ and $\gamma\gamma$. However, as
Ref.\cite{Martin:2008sv} shows, for a non-relativistic stoponium,
$\Gamma_{X\bar X}$ ($X = W, Z, h, t$) and $\Gamma_{\gamma\gamma}$ are
orders of magnitude smaller than $\Gamma_{gg}$. Thus, $Br(\eta_{\xt}
\to \gamma\gamma)$ remains essentially unchanged. Moreover, once
branching fractions for the cascades are included, the $X \bar X$
channels would remain undetectable.
And, of course, the dilepton
channel is even more suppressed. What remains irreconciled, though, is
the large width reported by ATLAS (in contrast to CMS). Even
accounting for detector effects and a low statistics, such a large
width can only be accommodated if the stoponium were to have a very
large decay width into (quasi-)invisible channels, much larger than
$\Gamma_{\xc\xc}$ (if kinematically allowed) is.

And, finally, we come to confirmatory tests for our hypothesis.
Obvious candidates are $\xt\xt^*$ production \cite{Kaufman:2015nda} in association with a light
object, such as a $\gamma, Z$, a jet, or given the large $A_t$, even
$h$. As already discussed, the
  strongest present bound for degenerate $\xt$-$\xc$ scenario comes
from $\xt\xt^*$+jets analysis and with more luminosity, other channels
become increasingly feasible. The characteristic feature of cMSSM with
degenerate $\xt$-$\xc$ is the existence of a heavy ($\sim 5m_{\xt}$,
see Fig.~\ref{param_space}) gluino decaying into top-stop
($t\xt^*+\bar t \xt$) pairs with 100\% BR. Therefore, $\gt \gt$-pair
production (although highly suppressed, see Table~\ref{cons}) gives
rise to confirmatory signatures namely, 2-boosted top-jets+large
$p_T\!\!\!\!\!\!/~$(because of large $m_{\gt}-m_{xt}$ splitting)
\cite{Ghosh:2012ud}, same-sign top-pairs+large $p_T\!\!\!\!\!\!/~$
(when both gluino decays into $t\xt^*$- or $\bar t \xt$-pairs)
\cite{Ajaib:2011hs} e.t.c., for our hypothesis. Assuming 80\%
efficiency ($\epsilon$) \cite{topjet_eff} for tagging hadronically
decaying boosted ($p^{(\rm top)}_T>300$ GeV) top at 13 TeV LHC, naive
estimation of $\sigma(\gt\gt)\times{\rm Br}\times{\epsilon}$ (see
Table~\ref{cons}) yields 45, 13 and 7 two high-$p_T$ top-jets + large
$p_T\!\!\!\!\!\!/~$ events for BP1, BP2 and BP3, respectively, for
${\cal L}=100~{\rm fb}^{-1}$. Therefore, probing boosted-tops
signature of BP1, BP2 and BP3 is possible with suitable cuts (large
$p_T$-cuts for top-jets and missing-$p_T$) to kill the SM background
completely without affecting the signal \cite{Ghosh:2012ud}. Same-sign
ditop signature of our hypothesis is particularly intriguing in view
of the reported excess of same-sign dilepton
(SSD) plus $b\bar b$ events at the LHC Run I~\cite{Aad:2015gdg}. 
The large
$\gt$-mass and thus, small $\sigma(\gt\gt)$ for the our BPs can not
explain the reported excess. However, were we deviate from the usual
cMSSM by incorporating non-universal boundary conditions for the
gaugino masses (as happens, for example, for non-trivial gauge kinetic
functions in the supergravity framework), the gluino can be
substantially lighter leading to a SSD+$b\bar b$ excess as reported by
ATLAS \cite{Aad:2015gdg}. And, finally, at a future $e^+e^-$ linear collider, a 
threshold scan would show up several of the stoponia, in particular 
the $nP$-states.

DC acknowledges partial support from the European Union's Horizon 2020
research and innovation programme under Marie Sklodowska-Curie grant
No 674896, and the R\&D grant of the University of Delhi. KG is
supported by DST (India) under INSPIRE Faculty Award.

\end{document}